\documentstyle[prd,aps,floats]{revtex}
\begin{document}
\draft

\input epsf \renewcommand{\topfraction}{0.8} 
\newcommand{\beq}{\begin{equation}}
\newcommand{\eeq}{\end{equation}}
\newcommand{\pbar}{\not{\!\partial}}
\newcommand{\dbar}{\not{\!{\!D}}}
\twocolumn[\hsize\textwidth\columnwidth\hsize\csname 
@twocolumnfalse\endcsname

\title{Is nonperturbative inflatino production during preheating a
real threat to cosmology? }   
\author{$^{\dagger}$Rouzbeh Allahverdi,~$^{\sharp}$Mar Bastero-Gil~and
~$^{*}$Anupam Mazumdar}    
\address{$^{\dagger}$ Physik Department, TU Muenchen, James Frank
Strasse, D-85748, Garching, 
Germany. \\
$^{\sharp}$ Scuola Normale Superiore, Piazza dei Cavalieri~7,
56126~Pisa, Italy \\ 
$^{*}$ ICTP, Strada Costiera~11, 34014~Trieste, Italy.}
\date{\today} 
\maketitle
\begin{abstract}
We discuss toy models where supersymmetry is broken due to
non-vanishing time-varying vacuum expectation value of the inflaton
field during preheating. We discuss the production of inflatino the
superpartner of inflaton due to vacuum fluctuations and then we argue
that they do not survive until nucleosynthesis and decay along with
the inflaton to produce a thermal bath after preheating. Thus the only
relevant remnant is the helicity $\pm 3/2$ gravitinos which can genuinely
cause problem to nucleosynthesis.
\end{abstract}

\pacs{PACS numbers: 98.80.Cq \hspace*{1.3cm} TUM-HEP-396/00/,
hep-ph/0012057} 

\vskip2pc]

%%%%%%%%%%%%%%%%%%%%%%%%%%%%%%%%%%%%%%%%%%%%%%%%%%%%%%%%%%%%%%%%%%%%%%%%
Inflation is perhaps one of the best paradigms of the early 
Universe which solves some of the nagging problems of the standard 
Big Bang cosmology \cite{guth}. One of the consequences of 
inflation is that it leaves the Universe extremely cold, virtually
devoid of entropy.  
Thus, the Universe requires to be reheated to a temperature at least more than 
${\cal O}(\rm MeV)$ to keep the successes of the Big Bang nucleosynthesis. 
Perhaps, one can imagine that the Universe reheats via the decay of the scalar 
field whose potential has dominated the Universe during the inflationary
regime. Inflation leaves the inflaton field extremely homogeneous 
except for the quantum fluctuations produced during inflation. The
perturbations  
keep their imprint intact to match the observed anisotropy in the present 
Universe which is one part in $10^{5}$ \cite{bunn}.
Once inflation ends, the mass of the inflaton field dominates over the Hubble
rate of expansion, and the  homogeneous inflaton field oscillates coherently 
around the bottom of the potential. If we assume  
chaotic inflation with a massive inflaton field, $m$, and potential 
$V= m^2\phi^2/2$,  
then, during the coherent oscillations the average pressure of the Universe 
within one Hubble time vanishes over many oscillations. As a result the 
decaying energy density of the Universe behaves as if it were in a matter 
dominated era with $\rho_{\phi}=\dot \phi^2/2+m^2\phi^2/2 \sim a^{-3}$, where 
$a$ is the scale factor of the expanding Universe. After couple of
oscillations the energy density in the scalar field redshifts away in
the same way as in the pressureless fluid but this does not lead to a
radiation dominated  Universe. To obtain a radiation dominated era,
the  inflaton field has to decay  
to other particles which will eventually lead to a thermalized plasma
with a finite temperature, usually known as the reheat temperature of
the Universe.  The inflaton decays when the Hubble parameter $H \sim
\Gamma_{\phi}$, where $\Gamma_{\phi}$ is the decay rate. The decay
rate essentially depends on the kind of couplings the inflaton has to
other particles \cite{dolgov}.  However, in between the end of inflation and
the beginning of the radiation era there can be an explosive
production of particles purely due to non-thermal effects. This new
wisdom has been realized in Refs.~\cite{brand}.  This is due to the
fact that the oscillations in the inflaton field are extremely coherent
and act as a Bose condensate fluid.  So, in principle, one can study
the quantum fluctuations of the inflaton quanta as well as bosonic and
fermionic fields which are coupled to the inflaton field via Yukawa,
gauge, or non-renormalizable couplings. Effectively, the problem
turns out to be quantizing the bosonic and the fermionic fields in a
time-varying inflaton background. This leads to an explosive
production of particles which does not depend on the background
temperature and it is purely an offshoot of a non-perturbative
analysis. The production of bosons and fermions differs in
its nature due to Pauli's exclusion principle, which prohibits
excessive production of fermions compared to their bosonic
counterparts \cite{ferm}. In this regard, recently it has been
realized that like fermions with spin $1/2$, other fermions with
higher spin can also be created from the vacuum fluctuations in a
time-varying scalar background.  In Ref.~\cite{anu}, the authors have
noticed that inspite of Planck mass suppressed couplings of spin $3/2$
particles to other fields, it is possible to excite them due to vacuum
fluctuations. This has lead to many consequences which we briefly
discuss in the next paragraph.

The spin $3/2$ gravitino occurs in supersymmetric theories as a
superpartner of the graviton. A massive spin $3/2$ has four helicity
states $\pm 3/2$ and $\pm 1/2$. A massless gravitinos only possess
$\pm 3/2$ helicity states. However, once supersymmetry is broken the
gravitinos become massive, and they possess all four helicity states.
In the early Universe supersymmetry can be broken due to non-zero
vacuum contribution of the inflaton energy density. If the inflaton
field is a scalar component of a chiral multiplet, then spontaneous 
supersymmetry breaking due to F-term leads to non-zero expectation
value of the fermionic field \cite{bailin}.
\begin{eqnarray}
\label{susyb}
\langle 0 |\delta_{\xi} \tilde \phi|0\rangle = \langle -i\pbar \phi
\xi - e^{G/2}G_{\phi}\xi \rangle \neq 0 \,,
\end{eqnarray}
where $\xi$ is the infinitesimal Grassmann-odd parameter, $\phi$ is
a scalar field responsible for inflation, $\tilde \phi$ is the
fermionic component of the inflaton in a single chiral field model,
which we may call inflatino, $G$ is the K\"ahler function defined
below, and $G_{\phi}$ is the derivative of the K\"ahler function with
respect to the inflaton field. The K\"ahler function is defined by
\begin{eqnarray}
G=\frac{\phi_{i} \phi^{i \ast}}{{\rm M}^2} + \ln\left(\frac{|W|}{{\rm
M}^3}\right)\,,
\end{eqnarray}
where we have assumed minimal K\"ahler function, $W$ is the
superpotential and ${\rm M} \equiv {\rm M}_{\rm p}/\sqrt{8\pi}$ is the
reduced Planck mass. Out of the two terms present in the right-hand
side of Eq.~(\ref{susyb}), we notice that the first term gives
a non-zero contribution during and after inflation, particularly
during preheating. Therefore, the dynamical effects of the inflaton field
breaks supersymmetry.

Soon after it has been realized that the helicity $\pm 3/2$ states of a
massive gravitino can be produced non-perturbatively \cite{anu}, it
has been shown that the helicity $\pm 1/2$ states of a massive gravitino
can also be produced from vacuum fluctuations \cite{kallosh}. However,
they are more abundantly produced compared to that of the helicity
$\pm 3/2$ states.  This can be easily understood in a simple way. For
the creation of particles from vacuum fluctuations, the adiabaticity
condition has to be broken which is usually measured by a rate of
change of a time-varying frequency of a given momentum mode. For
fermions the frequency depends on an effective mass parameter. For
example, for the helicity $\pm 3/2$, the mass parameter is essentially
Planck mass suppressed. It has been noticed in Refs.\cite{kallosh},
that the helicity $\pm 1/2$ states are massive due to the fact that
they eat the mass of the fermionic component of the inflaton.  This
statement is true for a single chiral case and it has been pointed out
that for helicity $\pm 1/2$ gravitinos, the adiabaticity condition is
broken much more strongly compared to that of helicity $\pm 3/2$
gravitinos \cite{giudice,maroto,mar,kallosh1}.

All these results obtained were interesting because gravitino plays a
key role in a standard Big Bang cosmology. If supersymmetry is
required to solve the gauge hierarchy problem, then, in the gravity
mediated supersymmetry breaking, the gravitino gets a mass around,
${\cal O}(\rm TeV)$. Since their couplings to other particles are
Planck mass suppressed, the life time of the gravitino at rest is
quite long, $\tau_{3/2} \sim M_{\rm p}^2/m_{3/2}^3 \sim
10^{5}(m_{3/2}/TeV)^{-3}$sec, \cite{cline}. We know that successful
nucleosynthesis depends on the baryon abundance: $Y_{\rm B}(T <{\rm MeV})
\equiv n_{3/2}/n_{\gamma}=10^{-10}$ \cite{sarkar}. The gravitino
decay products can easily change this ratio. Their decay products such
as gauge bosons and its gaugino partners, or high energy photons, can
generate a large entropy which will heat up the photons compared to
$\tau$ and $\mu$ neutrinos. The abundance of neutrinos essentially
determines the $^{4}{\rm He}$ abundance.  It was first pointed out in
Ref.~\cite{weinberg} that the gravitino mass must be larger than
$\sim 10$ TeV in order to keep the successes of the Big Bang
nucleosynthesis. On contrary, if the gravitinos were stable, and if
their mass exceeded $1$ KeV, they could easily overclose the Universe
in absence of inflation \cite{pagel}. However, after the end of
inflation the gravitinos can be produced from the thermal bath and
this constraints the temperature of the thermal bath in order not to
over produce them. At the time of nucleosynthesis the abundance is
given in terms of the reheat temperature, $Y_{3/2}(T<{\rm MeV}) \sim
10^{-2}(T_{\rm rh}/M_{\rm p})$ \cite{ellis}. Thus, we see that there
exists a strong constraint on the reheat temperature, $T_{\rm rh} \leq
10^{10}$ GeV, in order to maintain the baryon abundance one part in
$10^{10}$ during nucleosynthesis.  Since we know that 
non-perturbative creation of particles does not depend on 
temperature, it will be difficult to constrain a general parameter other
than the model parameters. Hence, this leads to a natural suspicion that
perhaps non-perturbative production of helicity $\pm 1/2$
gravitinos will cause a problem to nucleosynthesis.

The important point is that the inflaton has to completely decay
to give rise to a thermal bath with a reheat temperature at least more than a 
${\cal O}({\rm MeV})$, and the fermionic component of the inflaton, known as 
inflatino, inevitably decays along with the inflaton. We know during
the inflaton oscillations, the helicity $\pm 1/2$ states of the
gravitino eat the mass of the inflatino, 
and they essentially behave as an inflatino when the amplitude of the
inflaton oscillations has considerably dropped below ${M}_{P}$. As a result
they must also decay along with the inflaton. As we shall see that this
argument is quite robust and it should not depend if there were any
other source of supersymmetry breaking other than the inflaton sector.
During the preheating era of the Universe it is quite natural to think 
that the supersymmetry breaking due to the energy denisty stored in the
inflaton oscillations is far the most dominant source.

We will begin with an introduction of a supersymmetric inflationary
model with a single multiplet, and then we discuss decay rates of the
inflaton and the inflatino in two models: namely with Planck mass
suppressed coupling, and with Yukawa  couplings to the visible
sector.  We then establish an equivalence between the helicity $\pm
1/2$ gravitino interactions to its supercurrent to that of the
inflatino interactions in the supergravity Lagrangian when the
amplitude of the inflaton oscillations are small compared to the
Planck mass. In the last section we give a qualitative discussion upon
the gravitino decay when more than one chiral fields are present.

%%%%%%%%%%%%%%%%%%%%%%%%%%%%%%%%%%%%%%%%%%%%%%%%%%%%%%%%%%%%%%%%%%%%%%%%%%%%%%
%%%%%%%%%%%%%%%%%%%%%%%%%%%%%%%%%%%%%%%%%%%%%%%%%%%%%%%%%%%%%%%%%%%%%%%%%%%%%%

\section{Models With a Single Multiplet}

In the most part of this paper we shall focus on models where
supersymmtery is broken by a single multiplet and also responsible for
producing inflation in the early Universe. Nevertheless, to solve the
low-energy (i.e. electroweak scale) supersymmetry breaking we may
require some other sector, which can be a hidden sector, which we shall
not take into account here. In our case the source of a time-varying
supersymmetry breaking is the oscillations in the inflaton field
$\phi$. During these oscillations the fermionic partner of the
inflaton which we call here inflatino, whose mass is equal to the mass
of the inflaton, is eaten by the helicity $\pm 1/2$ components of the
gravitino to produce a massive gravitino. It has been suggested by many
authors in Refs.~\cite{giudice,maroto,kallosh1}, that in a limit when
$|\phi| < M_{\rm p}$, it is possible to use the inflatino mode
equation to study the behavior of the helicity $\pm 1/2$ states of the
gravitino. This can as well be understood from the point of view of an
Equivalence Theorem (ET), which demands that when the energy scale
${E} \gg m_{3/2}(t)$, the wavefunction of the gravitino for the
helicity $\pm 1/2$ components is approximately proportional to
$p_{\mu}/m_{3/2}(t)$, where $p_{\mu}$ is the momentum of the gravitino
and $m_{3/2}(t)$ is its time-varying effective mass. However, there is
a word of caution regarding the validity of ET in our 
calculation during the oscillations of the inflaton. In
principle, the time-varying mass of the gravitino can be larger than the
momentum during the oscillations, or perhaps $m_{3/2}(t) \propto
p_{\mu}$, and in both the cases ET cannot be trusted during the
inflaton oscillations. However, studying the inflatino mode equation
is not futile, because when the amplitude of the oscillations die down
due to the expansion of the Universe, it is possible to identify the
high momentum Fourier modes of the inflatino with those of the
helicity $\pm1/2$ gravitino. Therefore, only in those regions we can
identify the inflatino to the helicity $\pm 1/2$ gravitinos (in a
Fourier space), and we can therefore identify the Bogolyubov
coefficients which are related to the number density of the produced
helicity $\pm 1/2$ gravitinos. In this paper we are going to argue
that ET can also be used to study the decay of the helicity $\pm 1/2$
gravitinos. However, this means that by using ET we shall be able to
match the coupling strength of the helicity $\pm 1/2$ gravitinos to
that of the inflatinos. This we shall discuss in the coming sections.

%%%%%%%%%%%%%%%%%%%%%%%%%%%%%%%%%%%%%%%%%%%%%%%%%%%%%%%%%%%%%%%%%%%%%%%%%%%%%%%%%%%

\subsection{Inflaton decaying via gravitational coupling}

As a first example we consider a new inflation model proposed in
Ref.~\cite{ross}.  In this model the two distinct sectors are the
inflaton sector and the visible sector. These sectors interact with
each other only gravitationally, and can be considered separately in
the superpotential. The construction of the inflaton sector demands
supersymmetry is restored in the global minimum. While setting the
cosmological constant to zero, the simplest form of a superpotential
emerges \cite{ross}
\begin{eqnarray}
\label{superpotential}
{\rm I} = {\Delta^2 \over {\rm M}} (\Phi - {\rm M})^2\,,
\end{eqnarray}
where $\Delta$ determines the scale of inflation.  Here we have
denoted $\Phi$ as a superfield in the inflaton sector.  The amplitude
of the density perturbations produced during inflation by the
inflaton, $\phi$, is fixed by the COBE scale, which constraints
$\Delta/{\rm M} \approx 5\times 10^{-3}$. With this choice of
superpotential, inflation occurs for $\phi \ll {\rm M}$; the
oscillations take place around the minimum of potential ${\phi}_0 =
{\rm M}$, with a frequency $m_{\phi} \sim{{\Delta}^2/{\rm M}}$. The
scalar potential derived from the above superpotential has a form
\begin{eqnarray}
\label{potential}
V = e^{\sum_{j}\left(|\Phi_j|/{\rm M}\right)^2}\left(\sum_{k} \left |
\frac{\partial W_{\rm tot}}
{\partial\phi_k}+\frac{\phi^{\ast}_{k}W_{\rm tot}}{{\rm M}^2} \right
|^2 \, \right.  \nonumber \\ \left. -3\frac{|W_{\rm tot}|^2}{{\rm
M}^2}\right)\,,
\end{eqnarray}
where we have assumed minimal K\"ahler function and we consider the
total superpotential to be
\begin{eqnarray}
\label{totsuper}
W_{\rm tot}={\rm I}+{\rm L}\,,
\end{eqnarray}
where ${\rm L}$ can be recognized as a visible sector which contains
the light degrees of freedom. Before we begin our discussion on
decaying inflaton, we mention some of the essential points related to
this model. The dominant coupling of the inflaton to other light
degrees of freedom can be read from the potential,
Eq.~(\ref{potential}). Just by expanding the interference term in
Eq.~(\ref{potential}), we notice that the inflaton field can decay
only via trilinear coupling to the scalars. This certainly prevents
creation of such scalar fields via parametric resonance. Hence, the
decay of the inflaton is essentially perturbative in nature.  Under
the condition $m_{\phi}> H$, the decay rate of the inflaton does not
depend upon the curvature of the universe. However, the inflaton field
has a time varying amplitude, thus it is necessary to virialize
the mean value of the field. Otherwise, we may expand the inflaton
field around its minimum value ${\rm M}$, by assuming
\begin{eqnarray}
\label{shift}
\phi^{\prime}=\phi -{\rm M} -\hat{\phi} (t)\,,
\end{eqnarray}
where $\hat{\phi}$ is assumed to have pure oscillatory part with an
amplitude much less than one in units of reduced Planck mass. With the
help of Eq.~(\ref{shift}), it is easy to evaluate the interference
terms coming from the first squared term in the bracket in
Eq.~(\ref{potential}). The leading order term in the expansion
generates trilinear coupling to the matter sector from ${\rm L}$ with
a gravitational strength $\sim \Delta ^2/{\rm M}^2$, corresponding to
a decay width $\Gamma_{\phi} \sim {m}_{\phi}\left(\Delta ^2/{\rm
M}^2\right)^2$.  Since the mass of the inflaton is $m_{\phi}\sim
\Delta^2/{\rm M}$, this gives a finite decay width of the inflaton
\cite{ross}
\begin{eqnarray}
\label{decayrate}
\Gamma_{\phi} \sim \frac{\Delta^6}{{\rm M}^5}\,.
\end{eqnarray}
If we assume that the inflaton energy is converted into radiation
according to 
\begin{eqnarray}
\rho_{\phi} \approx \frac{\pi^2}{30}g_{\ast}T^4_{\rm r}\,,
\end{eqnarray}   
where $g_{\ast}$ is the relativistic degrees of freedom, then the
reheat temperature of the Universe can be estimated by
\begin{eqnarray}
T_{\rm r} \sim \left(\frac{30}{\pi^2
g_{\ast}}\right)^{1/4}\left(\Gamma_{\phi}{\rm M} \right)^{1/2} \approx
10^{-1} \frac{\Delta^3}{{\rm M}^2}\,.
\end{eqnarray}
For $\Delta/{\rm M} \sim 5 \times 10^{-3}$, the reheat temperature is
around $T_{\rm r} \sim 10^{8}$ GeV.

With this introduction we may now turn our attention to the decay of
the helicity $\pm1/2$ gravitinos which are created during the
oscillations of the inflaton field from the vacuum fluctuations. We
remind that gravitino production is completely non-thermal, and we
cannot associate their number density to any particular thermal
bath. We also notice that the mass of the gravitino need not necessarily
be that of a gravitino mass around ${\cal O}(\rm TeV)$. Especially, if
the inflaton sector has a supersymmetric preserving minimum with a
zero cosmological constant, then the mass of the gravitino vanishes
after the end of reheating, provided there is no other source of
supersymmetry breaking in any other sector.  However, the situation
can be little bit different if there exists other sources of supersymmetry
breaking. This we shall briefly discuss in the next section. 
In the present section we shall concentrate upon the former case 
where we argue that whatever helicity $\pm 1/2$ gravitinos are created during
preheating shall have to decay along with the inflaton to have a
successful Big Bang nucleosynthesis \footnote{ Here we would like to point
out that eventhough we are considering inflaton to be the only source of 
supersymmetry breaking which might seems unrealistic at first point but
the analysis is much simpler in this case and our arguments hold true 
even if there is another source of supersymmetry breaking. It is possible to
consider a Polnyi sector which breaks supersymmtry in the hidden sector
of the theory but as we shall see in the next section that its mass contribution
to the mass of the goldstino is very small and of the order of TeV. As a 
result the goldstino mass is essentially
dominated by the inflatino mass. In such a circumstance our 
present analysis of single chiral field is quite general and as long
as supersymmetry breaking due to the inflaton sector dominates
over any other sector which is quite normal to think,
we can apply our results faily well. All that we require 
is that the helicity $\pm 1/2$ gravitino mass is essentially being 
contributed by the inflatino mass.}.

The equation of motion for the helicity $\pm 1/2$ gravitino in a
cosmological background has been derived in the literature by using
alternative approaches \cite{kallosh,giudice,maroto,mar,kallosh1}.
The important realization is that when the amplitude of the
oscillations is much smaller than the reduced Planck mass, the
equation of motion for the helicity $\pm 1/2$ gravitino is effectively
that of the goldstino in a global supersymmetric limit. For a single
chiral case the goldstino is equivalent to the inflatino up to a phase.
Here we simply
quote some of the useful formulae which have already been established
in Refs.~\cite{kallosh,giudice,kallosh1}. The evolution of the
inflatino, which we define here as $\tilde{\phi}$, is given by
\cite{giudice}
\begin{eqnarray}
\label{goldstino}       
i{\gamma}^{0}\dot{\tilde{\phi}} - \hat{k}\tilde{\phi} - m_{\rm
eff}\tilde{\phi} = 0\,,
\end{eqnarray}
where $\hat{k} = {\gamma}^{i}{k}_{i}$, and ${k}_{i}$ are components of
the physical momentum, while $\gamma^{i}$ are the gamma matrices. The
validity of the above equation holds only in a global supersymmetric
limit.

When the amplitude of the inflaton oscillations $|\phi| \ll {\rm M}$,
the effective mass of the helicity $\pm 1/2$ gravitinos, for a single
chiral field and after phase rotation of the helicity $\pm 1/2$
gravitino field, is simply the mass of the fermionic component of the
inflaton field, which yields \cite{giudice}
\begin{eqnarray}
m_{\rm eff} =\frac{\partial ^2 {\rm I}}{\partial \phi^2}\,,
\end{eqnarray} 
where ${\rm I}$ is the inflaton superpotential. For a simple
superpotential Eq.~(\ref{superpotential}), the effective mass for the
helicity $\pm 1/2$ gravitino turns out to be equal to
\begin{eqnarray}
\label{m_1/2}
m_{\pm 1/2}\simeq m_{\rm eff} \sim \frac{\Delta ^2}{{\rm M}}\,,
\end{eqnarray}
which is the same as the mass of the inflaton \cite{ross}.  On the
other hand, for the same superpotential the other helicity $\pm 3/2$
gravitinos have mass given by \cite{bailin}
\begin{eqnarray}
\label{m_3/2}
m_{\pm 3/2} \equiv e^{\phi^2/2{\rm M}^2}\frac{|{\rm I}|}{{\rm M}^2}
\sim \frac{\Delta^2} {{\rm M}}\left(\frac{\phi(t)}{{\rm M}}\right)^2
\,,
\end{eqnarray} 
where we have assumed that the visible sector $\rm L$ does not
contribute to the gravitino mass.  This is quite apparent from the
above expressions Eqs.~(\ref{m_1/2},\ref{m_3/2}), that the mass of the
helicity $\pm 3/2$ gravitinos is not only suppressed by the reduced
Planck mass, but it also contains time varying amplitude of the
oscillations; $\sim \phi(t)$, which becomes vanishingly small near the
bottom of the potential. This is quite obvious because mass of the
helicity $\pm 3/2$ gravitinos is essentially generated by the dynamics
of the inflaton field, and it must vanish when supersymmetry is
restored at the global minimum of the potential.  Before we begin our
discussion on the decay of gravitinos, we compare different mass
scales with the Hubble expansion.  For the superpotential
Eq.~(\ref{superpotential}), the Hubble parameter it is given by $H
\sim (\Delta^2/{\rm M})(\phi(t)/{\rm M})$.  This leads to a simple
inequality in various mass scales which we must bear in mind
\begin{eqnarray}
\label{sca}
m_{\phi}\approx m_{\pm 1/2} > H > m_{\pm 3/2}\,.
\end{eqnarray} 

%%%%%%%%%%%%%%%%%%%%%%%%%%%%%%%%%%%%%%%%%%%%%%%%%%%%%%%%%%%%%%%%%%%%%%%%%%%%%%%
\subsection{Inflatino interactions}

In this subsection we analyze the decay rate of the inflatino. We
consider a following interaction which can be found in the matter
Lagrangian \cite{bailin}
\begin{eqnarray}
\label{dom}
|{\rm det}~e|^{-1}{\cal L}= -\frac{1}{2} e^{G/2} G^{i} G^{j} \bar
\chi_{i}{\chi}_{j{\rm L}} + {\rm h}.{\rm c}. \,,
\end{eqnarray}
where $G^{i}$ is the derivative of the K\"ahler potential with respect
to left and right chiral components. We can fix the index; $i= \phi$,
corresponding to the inflaton sector. This leaves the other index $j$
to run on the chiral components of the visible sector $\rm L$. Now
according to our previous discussion on the inflaton decay, here again,
we may assume that the inflatino is decaying into particles and
sparticles of the sector ${\rm L}$ with an interaction of the form
$\tilde \phi \chi_{j}\phi_{j}$, where $\tilde \phi$ is the inflatino,
$\chi_{j}$ is the fermionic component, and $\phi_{j}$ is the bosonic
fields belonging to the sector ${\rm L}$. The decay is essentially
mediated via gravitational coupling strength $\sim \Delta^2/{\rm
M}^2$. This corresponds to a decay width of the inflatino with mass
$\sim \Delta^2/{\rm M}$, which yields
\begin{eqnarray}
\Gamma_{\tilde \phi} \approx \frac{\Delta^6}{{\rm M}^5} \,.
\end{eqnarray}
This decay rate is the same as the decay rate of the inflaton.  The
result is not surprising because the inflatino mass is similar to the
inflaton mass, and both interact gravitationally to the visible sector
particles.  Now, if we argue that the helicity $\pm 1/2$ states of the
gravitino essentially behave as an inflatino in a global
supersymmetric limit, which corresponds to demanding that the
amplitude of the inflaton oscillations $|\phi | \ll {\rm M}$, then and
only then, we may argue that the helicity $\pm 1/2$ gravitinos decay
along with the inflaton into visible sector particles. This is an
important and generic conclusion which bears more attention. Our
statement is only true provided we believe in the equivalence between
inflatino and the helicity $\pm 1/2$ gravitinos at late stages of the
inflaton oscillations, i.e. when Eq.~(\ref{sca}) is satisfied.

Intuitively, our result makes sense, because if supersymmetry is
restored at the bottom of the potential in the absolute minimum, then,
only the $\pm 3/2$ components of the gravitino should survive, and not
the $\pm 1/2$ components of the gravitino. Thus, the helicity $\pm
1/2$ states must decay along with the inflaton decay. This situation
could have been different if there were a hidden sector, which was
responsible for supersymmetry breaking at an intermediate scale, which
would then communicated to the visible sector at low scale.  This we
shall discuss in the last section.  So far we have studied only the
inflatino interactions. However, to be more concrete we must study the
gravitino interactions which we shall discuss in the next subsection.

%%%%%%%%%%%%%%%%%%%%%%%%%%%%%%%%%%%%%%%%%%%%%%%%%%%%%%%%%%%%%%%%%%%%%%%%%%%%%%%%%%%%%%%%%%%

\subsection{Interactions of the gravitino}

The gravitino interaction terms appear from the couplings between the
gravitino field and the supercurrent
\begin{eqnarray}
\label{lag1}
{\cal L}_{\psi J}=\frac{1}{\sqrt{2}{\rm M}}\bar \Psi_{\mu}\dbar
 \phi^{\ast j} \gamma^{\mu}\chi_{j {\rm L}} +\frac{i}{\sqrt{2}{\rm
 M}}e^{G/2}G^{i}\bar \Psi_{\mu}\gamma^{\mu}\chi_{i {\rm L}} \nonumber
 \\ + {\rm h.c.} \,,
\end{eqnarray}
where $\mu$ stands for the space-time index, $\chi_{i}$ is a fermionic
field and $\phi^{i}$ is a bosonic field. Here the subscripts $i,j$
correspond to the visible sector ${\rm L}$, which contains the light
degrees of freedom.  We have neglected the vector multiplets in the
above equation and assumed $\phi$ to be homogeneous. The total
derivative $D_{\mu}$is defined by
\begin{eqnarray}
D_{\mu} = \partial_{\mu} + \frac{1}{2} \omega_{\mu ab}\sigma^{ab}\,,
\end{eqnarray}
where $\omega_{\mu ab}$ is the spin connection.

It is to be mentioned that apart from the derivative coupling of the
chiral field to the gravitino, we have an extra interaction term which
is not usually considered otherwise. In fact the interaction terms
proportional to $\gamma^{\mu}\Psi_{\mu}$ are usually not necessary in
a static limit of the background field (i.e. inflaton field ), because
$\gamma^{\mu}\Psi_{\mu}=0$ acts as a constraint for a gravitino field
in a static background. However, this need not be true in a non-static
background. It has been shown that in an expanding Universe, and in a
time-varying inflaton background, $\pm 1/2$ helicity states follow
$\gamma_{\mu}\Psi^{\mu}\neq 0$ \cite{kallosh}. Although, the same
constraint continues to hold good for the helicity $\pm 3/2$
components of the gravitino in the same background along with the
Dirac equation \cite{anu}.  Thus, both the terms in Eq.~(\ref{lag1})
should be taken into account to study the efficient decay of the
gravitino.  In this subsection we will study the decay by assuming the
validity of the equivalence between the helicity $\pm 1/2$ states of
the gravitino and that of the inflatino at late stages of
oscillations, i.e. when Eq.~(\ref{sca}) is satisfied.

After several oscillations of the inflaton field $|\phi| \ll {\rm M}$,
or, equivalently $H \ll m$. Note that under this condition, the kinematics
of the inflaton, such as decay rate does not depend on the curvature of
the Universe. As a result 
of this the decay rate of the inflaton coicides with that of the flat
space-time limit. All the fields whose effective mass is larger than 
the Hubble parameter during the oscillations of the inflaton would 
actually not feel any effect of curvature of the Universe. 
Since, the effective mass of the helicity $\pm 1/2$ gravitino is
similar to the mass of the inflaton, and, it is much larger than the
Hubble parameter, suggests that we can study their evolution by
neglecting the curvature of the Universe. 
Therefore, we replace $\pm 1/2$ helicity of the gravitino by an ansatz
\begin{eqnarray}
\label{ansatz}
\Psi_{\mu} \sim  \sqrt{\frac{2}{3}}\frac{\rm
M}{\rho_{\phi}^{1/2}}{\partial}_{\mu}\eta \,,
\end{eqnarray}
where $\eta$ represents the spin $1/2$ fermionic field, which we shall
interpret as a goldstino instead of inflatino. At this moment this
prescription seems to be unwarranted, but, we shall see that this choice
of derivative wavefunction leads to the interactions of the helicity
$\pm 1/2$ gravitino to that of the inflatino. This prescription has
also been used in Ref.~\cite{giudice}. The goldstino is
however related to the inflatino by a phase factor, and, it is
expressed in Eq.~(\ref{app3}). The above expression is exactly 
the wavefunction of the helicity $\pm 1/2$ gravitino in terms of
goldstino in the limit when the energy scale of the gravitino
is larger than its effective mass. This expression has been 
previously used in Refs.~\cite{fayet,gherghetta,moroi}, where 
the authors have been studying the scattering processes of the 
helicity $\pm 1/2$ gravitino in a limiting case when the energy
scale of the gravitino is larger than its mass in a flat space-time.
Here, again we have a similar situation where the helicity $\pm 1/2$
gravitino does not feel the Hubble expansion, however, the only 
difference is that now supersymmetry is broken due to the oscillating
scalar field rather than the static vacuum contribution. This 
obviously makes the difference. Notice,
that instead of the mass of the gravitino $m_{3/2} \sim 1$TeV, now 
we have $\rho_{\phi}^{1/2}/M$, this is precisly because of the reason
we have mentioned above, here supersymmetry is effectively broken 
due to the presence of the finite energy contribution of the inflaton field.
The oscillations of the inflaton field also ensures that one has to 
include the interactions which are also proportional to 
$\gamma^{\mu}\Psi_{\mu}$. Another fact that the
equation of motion of helicity $\pm 1/2$ gravitino is the same as that
of the goldstino for $H \ll m$, as indicated in Refs.~\cite{kallosh,giudice},
further supports our approximation. We reiterate that we shall always work 
in a limit $\phi/{\rm M} \ll 1$.

Substituting Eq.~(\ref{ansatz}), in Eq.~(\ref{lag1}), we derive an
effective Lagrangian. Which yields
\begin{eqnarray}
\label{siml}
{\cal L}_{\rm eff}=\frac{1}{\sqrt{3}\rho_{\phi}^{1/2}}\partial_{\nu}
{\varphi ^{\ast}} \partial_{\mu}{\bar\eta} {\gamma}^{\nu} {\gamma}^{\mu}
\left(\frac{1+\gamma_{5}}{2}\right)\chi \, \nonumber \\
+\frac{i}{\sqrt{3}\rho_{\phi}^{1/2}}m_{\chi}\varphi^{\ast}
\partial_{\mu}\bar\eta \gamma^{\mu}\left( \frac{1+\gamma_{5}}{2}\right)\chi
+\rm{h.c.}\,,
\end{eqnarray}
where $\varphi$ denotes the bosonic field in a visible sector $\rm L$,
the spinor component is defined by $\chi$; its mass can be written
as
\begin{eqnarray}
\label{massl}
m_{\chi} \approx e^{\phi^2/2{\rm M}^2}\frac{\partial^2 {\rm
L}}{\partial\varphi^2} \,.
\end{eqnarray}
We will use $m_{\chi}$ quite often to compactify our notations. In
fact the mass of the fermion need not be a constant and can have field
dependency. We can simplify Eq.~(\ref{siml}) if we follow the below
mentioned identities for the Majorana spinors
\begin{eqnarray} 
\label{id1}
\bar{\eta} {\gamma}^{\mu} \chi &=& - \bar{\chi} {\gamma}^{\mu} \eta\,,
\nonumber \\ \bar{\eta} {\gamma}^{\mu}{\gamma}^{5} \chi &=& \bar{\chi}
{\gamma}^{\mu} {\gamma}^{5} \eta \,,\nonumber \\ \bar{\eta}
{\gamma}_{5} \chi &=& \bar{\chi} {\gamma}_{5} \eta \,.
\end{eqnarray}
With the help of Eq.~(\ref{id1}), we can derive an effective
Lagrangian after some algebraic manipulation, which reads
\begin{eqnarray}
\label{app2}
{\cal L}_{\rm eff} &=& \frac{1}{\sqrt{3}\rho_{\phi}^{1/2}} \left[
\left(m^2_{\varphi \rm R}-m^2_{\chi}\right) \varphi_{\rm R}\bar \eta
\chi -i\frac{\partial m_{\chi}}{\partial t}\varphi_{\rm R} \bar \eta
\gamma^{0}\chi\right]\, \nonumber \\ &&-
\frac{i}{\sqrt{3}\rho_{\phi}^{1/2}} \left[ \left(m^2_{\varphi \rm
I}-m^2_{\chi}\right) \varphi_{\rm I}\bar \eta {\gamma}_{5} \chi +
i\frac{\partial m_{\chi}}{\partial t}\varphi_{\rm I} \bar \eta
\gamma^{0} {\gamma}_{5} \chi \right]\, \nonumber \\ &+&{\rm h.c.} +
{\rm total~ derivative}\,,
\end{eqnarray}
where $m_{\varphi \rm R}$ denotes the real part of the light bosonic
field $\varphi_{\rm R}$ residing in the sector ${\rm L}$. While
deriving the above expression we have neglected the time derivative of
the energy density.  Eq.~(\ref{app2}) can be further simplified if we
assume that the mass splitting between $\varphi$ and $\chi$ is due to
supersymmetry breaking by the inflaton oscillations. To simplify the
situation we will be assuming that the visible sector must contain the
quadratic terms in the superpotential.  This can be written as
\begin{eqnarray}
\label{app33} 
{\rm L} = \frac{1}{2}\frac{\partial^2 {\rm L}}{\partial \Lambda^2}\Lambda^2
+ ...\,,
\end{eqnarray}
where $\Lambda(\varphi,\chi)$ denotes the superfield, and $...$ terms
can contribute due to other possibilities in the superpotential, which
we shall not take into account here. Now we will explicitly show that
if the inflaton sector and the visible sector interacts
gravitationally, then it is possible to derive an effective inflatino
Lagrangian which will have a similar coupling to Eq.~(\ref{dom}). To
get the desired result we first need to know the mass splitting
between the fields of the visible sector. To get the mass splitting we
expand Eq.~(\ref{potential}), with the help of Eqs.~(\ref{totsuper})
and (\ref{app33}), while considering only the dominant terms in the
potential which is due to the interference terms. We also take help of
Eq.~(\ref{massl}) to obtain
\begin{eqnarray}
\label{split}
m^2_{\varphi \rm R}-m^2_{\chi} \approx e^{\phi^2/2{\rm
M}^2}\langle\frac{\phi}{{\rm M}}\rangle \langle \frac{1}{{\rm
M}}\frac{\partial {\rm I}}{\partial \phi}\rangle m_{\chi}\,,
\end{eqnarray}
and, similar expression holds for $m^2_{\varphi \rm I}-m^2_{\chi}$,
except for the negative sign on the right hand side. Similarly, one
may also obtain
\begin{eqnarray}
\label{appen}
\frac{\partial{{m}_{\chi}}}{\partial{t}} \approx \langle \frac{\dot \phi}{{\rm M}}\rangle
\langle \frac{\phi}{{\rm M}}\rangle m_{\chi}\,.
\end{eqnarray}
We also notice that the goldstino can be expressed as in
Refs.~\cite{giudice,mar,kallosh1}
\begin{eqnarray}
\label{app3}
\eta =\frac{1}{\rho_{\phi}^{1/2}}\left(i\gamma^{0}\frac{\partial
\phi}{\partial t}- e^{\phi^2/2{\rm M}^2}\frac{\partial {\rm
I}}{\partial \phi}\right)\tilde \phi\,,
\end{eqnarray}
where inflatino is represented by $\tilde \phi$.  Here we have
explicitly used the fact that the dynamics of the inflaton is breaking
supersymmetry. In Eq.~(\ref{app3}), we have only retained the leading
order terms and neglected ${\cal O}(1/{\rm M}^2)$ terms. Now, with the
help of Eqs.~(\ref{split}, \ref{appen},\ref{app3}), we simplify
Eq.~(\ref{app2})
\begin{eqnarray}
\label{app4}
{\cal L}_{\rm eff}={m}_{\chi} {\phi \over {\rm M}^2} {\varphi}_{R}
\bar{\tilde \phi} \chi + i {m}_{\chi} {\phi \over {\rm M}^2}
{\varphi}_{I} \bar{\tilde \phi} {\gamma}_{5} \chi + {\rm h.c.}\,,
\end{eqnarray} 
where we have used the identity ${{\gamma}^{0 \dagger}} =
{\gamma}^{0}$, and the fact that the following relation holds
\begin{eqnarray}
\label{relation}
\left (-i {\gamma}^{0}{d \phi \over dt} - e^{\phi^2/2{\rm M}^2}
{\partial{{\rm I}} \over \partial{\phi}}\right)\left(e^{\phi^2/2{\rm
M}^2} {\partial{{\rm I}} \over \partial{\phi}}- i {\gamma}^{0}{d \phi
\over dt}\right)\, \nonumber \\ =- {\rho}_{\phi} \,.
\end{eqnarray}
Now it is interesting to note that Eq.~(\ref{app4}), upto a leading
order actually leads to a familiar form
\begin{eqnarray}
\label{app5}
{\cal L}_{\rm eff} \approx e^{G/2} {\partial{G} \over \partial{\phi}}
{\partial{G} \over \partial{\varphi}} \bar{\tilde \phi} \chi_{\rm L}
+{\rm h.c.} \,,
\end{eqnarray}
which is the inflatino coupling in Eq.~(\ref{dom}).  While deriving
the above expression we have assumed Eq.~(\ref{massl}).  This clearly
indicates that at late time of the inflaton oscillations when we
recognize the helicity $\pm 1/2$ component of the gravitino as a
goldstino, we essentially get similar coupling to the visible sector
as that of the inflatino.  This is the most important equivalence
which establishes the fact that, since, for any successful
inflationary model inflaton has to decay, and, so does the inflatino,
the helicity $\pm 1/2$ component of the gravitino must also decay if
the inflaton oscillations is the only viable source of supersymmetry
breaking at that time.  Our result is strictly correct for a single
chiral field responsible for supersymmetry breaking. A further
generalization to multi-chiral field supersymmetry breaking is more
involved and we leave this for our future investigation.

Now, we move onto a toy model where the inflaton sector and the
visible sectors are coupled via Yukawa couplings. We will establish
similar result as we have already obtained in this section.
%%%%%%%%%%%%%%%%%%%%%%%%%%%%%%%%%%%%%%%%%%%%%%%%%%%%%%%%%%%%%%%%%%%%%%%%%%%%%%%%%%%%%%%%%%
%%%%%%%%%%%%%%%%%%%%%%%%%%%%%%%%%%%%%%%%%%%%%%%%%%%%%%%%%%%%%%%%%%%%%%%%%%%%%%%%%%%%%%%%%%

\subsection{Model with a Yukawa coupling to the Inflaton}

As a second example we consider a model with a following
superpotential
\begin{eqnarray}
\label{superpot2}
W = {1 \over 2} m {\Phi}^2 + \frac{1}{2}h\Phi \Sigma^2 \,,
\end{eqnarray}
where $\Phi$ contains the inflaton field, which is responsible for the
slow-roll inflation. However, now the inflaton field has an explicit
Yukawa coupling to the matter sector given by the second term in
Eq.~(\ref{superpot2}).  Such a coupling will enable inflaton to decay
much more efficiently. Such a superpotential leads to interaction
terms $hm\phi \sigma \sigma$, $h\phi \tilde \sigma \tilde \sigma$,
$h\tilde\phi \tilde \sigma \sigma$, where $\phi $ is the inflaton
field, $\tilde \phi$ is the inflatino, $\sigma $ is a light bosonic
field, and its fermionic partner has been denoted by $\tilde \sigma$.
The estimated rate of the inflaton decaying to fermionic component
$\tilde \sigma~\tilde \sigma$ is given by $\Gamma_{\phi} \sim
(h^2/8\pi)m$.

In general the Yukawa coupling between $\Phi$ and $\Sigma$ multiplets
can also result in the oscillations along the $\sigma$ field. If the
$\sigma $ field eventually decays into other products much before the
oscillations in $\sigma$ commences, then, it can still be a viable
model to imagine that supersymmetry is broken by the inflaton
field only. But in general, this may lead to a more complicated
situation where supersymmtery is broken by several multiplets.  However, it is
possible to prevent this provided we require that the $\phi$-induced
mass to the $\sigma$ field is much smaller than the Hubble expansion,
i.e. $h \phi < H$, which implies $h <m/{\rm M}$. We note that this
will also insure that $\sigma$ and 
$\tilde{\sigma}$ are not produced via parametric resonance.  A viable
choice of parameters which can lead to an inflationary
paradigm for the $\phi$ field in a quadratic potential are; $m =
10^{13}$ GeV, and, a small Yukawa coupling $h = 10^{-7}$, which
ensures that at late stages of the 
inflaton oscillations, ${\phi /{\rm M}} \leq 10^{-14}$, the inflaton
is decaying perturbatively.  Following our previous discussion, again,
we argue here that since the inflatino mass is same as that of the
mass of the inflaton, and, if the helicity $\pm 1/2$ components of the
gravitino is recognized as inflatino at late stages of the inflaton
oscillations, then they must decay to $\sigma $, or $\tilde \sigma $
via a Yukawa coupling.

So far, we have been looking upon direct inflatino coupling to
$\sigma$ and $\tilde \sigma$.  However, we may now repeat the same
analysis as we have shown earlier that indeed the helicity $\pm 1/2$
components of the gravitino has a similar coupling as that of the
inflatino by using the equivalence theorem. The generalization is
quite simple and we recognize that
\begin{eqnarray}
{{\partial}^2 W \over {\partial{\Sigma}}^2} = h \Phi \,,
\end{eqnarray}
where $\Sigma$ and $\Phi$ are the superfields denoted in
Eq.~(\ref{superpot2}).  The mass of the fermion in this model is given
by ${m}_{\tilde \sigma} \approx e^{\phi^2 /2{\rm M}^2}h\langle \phi
\rangle$. Following our earlier argument we can find out the leading
order contribution to the mass splitting, which yields
\begin{eqnarray}
\label{app6}
m^2_{\sigma \rm R} - m^2_{\tilde\sigma} \approx h\langle \frac{\partial {\rm
I}} {\partial \phi}\rangle\,.
\end{eqnarray}
where ${\rm I} \equiv (1/2)m\Phi^2$.  Similarly, one can also derive
an expression for $m^2_{\sigma \rm I} - m^2_{\tilde\sigma}$, which differs from
the above by a negative sign. While deriving Eq.~(\ref{app6}), we have
neglected the Planck mass suppressed contributions which would anyhow
be insignificant at late times. The analogue of Eq.~(\ref{appen}) can
be expressed as
\begin{eqnarray}
\label{app7}
{\partial{{m}_{\tilde\sigma}} \over \partial{t}} \approx h \langle \frac{\partial \phi}{\partial t}
\rangle \,,
\end{eqnarray}
and, one can also derive an effective Lagrangian with the help of
Eqs.~(\ref{app2}), (\ref{relation}), (\ref{app6}) and
(\ref{app7}). Which yields
\begin{eqnarray}
\label{app8}
{\cal L}_{\rm eff} \sim h ({\sigma}_{\rm R} \bar{\tilde \phi} \tilde\sigma +
i {\sigma}_{\rm I} \bar{\tilde \phi}{\gamma}_{5} \tilde\sigma) + {\rm h.c.}
\,,
\end{eqnarray}
where inflatino is denoted by $\tilde \phi$. After some calculation it
can be shown that Eq.~(\ref{app8}), actually leads to an expression
\begin{eqnarray}
\label{app9}
{\cal L}_{\rm eff} \sim h {\sigma}^{*}\bar{\tilde\phi}\tilde\sigma_{\rm R}
+ {\rm h.c.}\,.
\end{eqnarray}
This reinsures our earlier claim that the equivalence between the
helicity $\pm 1/2$ gravitino and the goldstino is viable at late times
of the inflaton oscillations. This equivalence is not only important
for studying the production of the helicity $\pm 1/2$
components of the gravitino, but also describing the decay of the
helicity $\pm 1/2$ gravitinos.

So far, we have not spoken any word on the other helicity states of
the gravitino, namely $\pm 3/2$. The reason is it is extremely
difficult to study their decay, precisely because the mass of the
helicity $\pm 3/2$ is solely due to the dynamics of the inflaton field
\cite{anu}. Their effective mass is Planck mass suppressed, and also
depends on the amplitude of the oscillations of the inflaton
field. This leads to an obvious result that if there is no other
source of supersymmetry breaking other than the inflaton oscillations,
then, the effective mass for the helicity $\pm 3/2$ component should
vanish at the end of reheating. It is difficult to make a precise
calculation for the decay of the helicity $\pm 3/2$ gravitinos.
However, we believe that their survival does not depend on the
inflaton decay as they have no goldstino nature. 
Next, we discuss
qualitatively what would happen if the hidden sector supersymmetry
breaking is also taken into account.

%%%%%%%%%%%%%%%%%%%%%%%%%%%%%%%%%%%%%%%%%%%%%%%%%%%%%%%%%%%%%%%%%%%%%%%%%%%%%%%%%%%%%%%%%
%%%%%%%%%%%%%%%%%%%%%%%%%%%%%%%%%%%%%%%%%%%%%%%%%%%%%%%%%%%%%%%%%%%%%%%%%%%%%%%%%%%%%%%%%%   
\section{Models With Several Multiplets}

Once we invoke more than one sectors, and treat them at equal level,
the problem of gravitino production becomes more complicated. This
problem has been addressed in Refs.~\cite{giudice,mar,kallosh1} to
some extent, and yet lot to be understood in this direction. In this
case it has been realized that the goldstino is a linear combination
of all the fermions, and as a result, even if we use the
goldstino-gravitino equivalence we cannot in general guarantee that a
major contribution to the goldstino mass is coming from the fermionic
component of the inflaton. There are some interesting cases where the
multi field case can be expressed as a single field, such as
supersymmetric hybrid inflation model where effectively the two fields
behave as if there were a single degree of freedom \cite{mar}.  In
such a model it is possible to extract the goldstino mass, which is
again of the order of the inflaton mass. One can then discuss the
decay rate of the inflatino in this model \cite{mar}, and the
inflatino decay rate to the light degrees of freedom would exactly be
the same as that of the inflaton.  Interesting question would be to
address a problem where there exists a hidden sector which is
responsible for supersymmetry breaking in that sector, and also
responsible for mediating supersymmetry breaking gravitationally to
the observable sector. In such a case the gravitino will have an
effective mass $\sim {\cal O}({\rm TeV})$ at a low energy scale.  
So, keeping this in mind we may consider a simple toy model with a 
following superpotential
\begin{eqnarray}
\label{newsuperpot}
W = {1 \over 2} m_{1} {\Phi}^2 + m_{2}^2[Z+ (2 - \sqrt{3}){\rm M}]\,,
\end{eqnarray}
where $\Phi$ and $Z$ are inflaton and Polonyi multiplets respectively.
We assume that $\phi$ field is responsible for inflation, so we set
$m_{1} = 10^{13}$ GeV to produce adequate density perturbation, while
setting $m_{2} = 10^{11}$ GeV for giving an effective mass to the
gravitino around ${\cal O}({\rm TeV})$. An interesting discussion
regarding this model has been sketched in Ref.~\cite{kallosh1}.

A serious difficulty which immediately arises is that one derives a
set of coupled equations for the helicity $\pm 1/2$ gravitino and
other fermionic degrees of freedom \cite{giudice,kallosh1}. It has
been shown in Ref.~\cite{kallosh1}, that in a global supersymmetric
limit, this set of equations is reduced to a coupled set of equations
for the goldstino and the transverse combination of the fermions. This
suggests that there exists a mixing between the goldstino and the
transverse combination of the fermions. As a result one cannot
describe the goldstino in a mass eigenstate, and thus, it is also
difficult to estimate the evolution of their number densities. There
are many technical difficulties because there are essentially two time
scales in the problem.  The first one is related to the fact that the
effective mass scale of the bosons oscillating and exciting the
fermionic modes, and, the other one is related to the mixing between
the goldstino and the transverse combination of the fermions (for
details, we refer the readers Refs.~\cite{giudice,kallosh1}).  In
general one can derive a relationship between the two time scales, but
this is a non-trivial task and we do not have enough tools to address
this problem.

For the above superpotential Eq.~(\ref{newsuperpot}), the inflaton and
the Polonyi sectors have only gravitational interactions. The
fermionic components $\tilde{\phi}$ and $\tilde{z}$ have masses
$m_{1}$ and zero respectively in the global supersymmetric limit.  The
goldstino in this model is a linear combination of the fermionic
components from both the sectors. As long as the energy density is
dominated by the inflaton field, the helicity $\pm1/2$ gravitinos 
essentially behave as an inflatino, because the mass contribution
to the goldstino from the Polnyi sector is much smaller 
$\sim {\cal O}(\rm TeV)$. This 
particular case is quite interesting and we can analyse the decay 
of the gravitino by assuming that the gravitinos are created from
the vacuum fluctuations due to the inflaton oscillations, whose 
energy density is dominating the Universe. The helicity $\pm 1/2$
gravitinos produced during preheating will essentially decay because
they are essentially the inflatino components and so their couplings
are determined in the same fashion as that of the inflaton.

However, the energy density in the inflaton sector is decreasing 
in time, and, when the Hubble expansion $\sim H < {\cal O}(\rm TeV)$, the
$\tilde{z}$ component dominates the goldstino. Usually, the mixing between
the inflatino and $\tilde{z}$ is minimal and Planck mass suppressed, so,
the fermions which are produced during preheating will decay again in 
the form of inflatino and cause no trouble for nucleosynthsis, yet there
is a finite probability to mix the fermionic states and conversion of
inflatino to the fermionic partner of the Polonyi field. Though, we 
shall not discuss this possibility in this paper.  
One can also imagine that the oscillations in
the Polonyi sector are also induced at $H \approx {\cal O}(\rm TeV)$. 
Once, $z$ field starts oscillating, one might expect supersymmetry 
is broken by the oscillations in $z$ direction also, and, as a 
result gravitinos can as well be excited. One may also suspect 
that the late production of the helicity $\pm 1/2$ gravitinos will 
dominate and the problem of gravitino decay still persists. The suspicion is
not fully correct because the number density of the helicities $\pm 1/2$
and $\pm 3/2$ are more or less equal now. This is because superpotential
contribution to the mass of the
fermionic component of the Polonyi field is very small 
$\sim {\cal O}(\rm TeV)$ and the only
time-varying scale is due to time-varying mass of the gravitino $\sim
e^{zz^{\ast}/2{\rm M}^2}|W|/{\rm M}^2$. The presence of the Planck
mass suppression prohibits explosive production of the gravitinos at 
late times, so especially in the model we have considered, the late time
production of helicity $\pm 1/2$ can not be very abundant.
But, now the problem could be much more severe, because these gravitinos with
both the helicities are produced much later, and their effective
masses are also very small roughly of the order of TeV. This leads to
extremely slow decay rate of these gravitinos which may cause a
problem to the Big Bang nucleosynthesis. This picture is similar to 
late production of gravitinos discussed in Refs.~\cite{lyth}. 
Furthermore, the oscillating Polonyi field leads to an even more 
serious problem, i.e. the moduli problem, of which there is no 
satisfactory way out.

Finally, we mention and also pointed out in Ref.~\cite{kallosh1}, that 
if the fermionic components mix freely, the inflatinos can be  
converted to $\tilde{z}$ (which is the field eventually eaten by the
gravitino).
This presumably occurs around the time when contributions to
supersymmtery breaking from the inflation sector and the Polonyi
sector become comparable. This problem is
analogue to the neutrino flavor conversion and the relevant question
is to ask the conversion probability. 
As mentioned, we beileve that an effcicient conversion will not take
place for the Polonyi model. An efficient conversion nevertheless results
in a large abundance (i.e. comparable to the abundances which are
produced during preheating) of $\tilde{z}$ fermion, on top of what
is produced due to oscillations of the Polonyi field\footnote{This is
the abundance of $\tilde{z}$ fermion which will eventually
determine the abundance of helicity $\pm 1/2$ gravitinos.}. We notice that if the inflatino decays before 
$H \approx {\cal O}({\rm TeV})$, then the abundance of the 
inflatinos prior to conversion will decrease leading to a smaller 
abundance for $\tilde{z}$ (and consequently helicity $\pm 1/2$
gravitinos) even after an efficient conversion. The
quantitative analysis 
is beyond the scope of this paper and we leave that
for future investigation.

%%%%%%%%%%%%%%%%%%%%%%%%%%%%%%%%%%%%%%%%%%%%%%%%%%%%%%%%%%%%%%%%%%%%%%%%%%%%%%%%%%%%%%%%%%

\section{conclusion}

Our main result of this paper is to show that the inflatino coupling
to the matter field is similar to that of the helicity $\pm 1/2$
gravitinos. This merely confirms that the gravitino interaction with 
the supercurrent actually leads to the same interactions as that of 
the inflatinos when the amplitude of the inflaton oscillations is 
small $|\phi| \ll {\rm M}$, under the assumption that the helicity 
$\pm 1/2$ component of the gravitino behaves as a goldstino for a momentum 
larger than the gravitino mass in a time-varying background.  
Then we have argued that the 
production of helicity $\pm 1/2$ states of the gravitino, especially
for models where supersymmetry breaking scale is dominated by the 
inflaton energy scale, cannot be considered as a threat for nucleosynthesis. 
Their overproduction can be easily understood from the presence of a 
second derivative of the superpotential with respect to the super 
fields in the equation of motion for the helicity $\pm 1/2$ gravitinos. 
This gives rise to an effective mass for the helicity $\pm 1/2$ 
gravitinos, which is equivalent to the mass of the fermionic component 
of the inflaton, known as inflatino. The statement is true only if the 
inflaton sector has a single multiplet. In some sense helicity 
$\pm 1/2$ states eat the mass of the goldstino, which is related to 
the inflatino by an appropriate phase.  These states remember their 
goldstino nature and this is the reason why they are produced very 
efficiently compared to the helicity $\pm 3/2$ states. In this paper 
we have argued that the same goldstino nature come into rescue the late decay
of the helicity $\pm 1/2$ gravitino.  It has been argued by many
authors that the helicity $\pm 1/2$ gravitinos effectively behave like
goldstino just after couple of inflaton oscillations.  This, together with a
requirement that the inflaton must decay to give a successful
nucleosynthesis, leads to an efficient decay of the goldstino, or, the
helicity $\pm 1/2$ gravitinos.  Thus, they must not survive until
nucleosynthesis, and hence they should not be considered as a threat
to nucleosynthesis. This argument holds perfectly well for a single
chiral field where the goldstino is inflatino with some additional
phase. However, the extension of this argument in some models 
where there are more than one sectors of supersymetry breaking can be
made applicable, provided supersymmetry breaking scale is still dominated
by the inflaton energy density. Such a situation can arise if there 
exists a Polonyi field in the hidden sector, which we have briefly discussed.
However, we still lack a complete formal tools to explore all possibilities
such as mixing between the fermionic components of the inflaton sector and the
Polonyi sector. This can in principle change the abundance of the helicity 
$\pm 1/2$ component of the gravitinos and a detailed study is certainly 
required in this direction.

It is important to note that the above 
discussion does not apply to the helicity $\pm 3/2$ gravitinos.  
The production of these states during preheating is always Planck mass 
suppressed and their existence is also independent of the goldstino, 
so they decay quite late.  Due to time varying nature of their masses 
it is always hard to estimate their decay rate. It is also true that  
the helicity $\pm 3/2$ states are in general produced in less abundance 
than helicity $\pm 1/2$ states, however, their abundance cannot be 
neglected as pointed out in Refs.~\cite{anu,mar}. 
For a single multiplet they are the only genuine
threat to Big Bang nucleosynthesis.

%%%%%%%%%%%%%%%%%%%%%%%%%%%%%%%%%%%%%%%%%%%%%%%%%%%%%%%%%%%%%%%%%%%%%%%%%%%%%%%%%%%%%%%%%%%

%%%%%%%%%%%%%%%%%%%%%%%%%%%%%%%%%%%%%%%%%%%%%%%%%%%%%%%%%%%%%%%%%%%%%%%%%%%%%%%%%%%%%%%%
\section*{Acknowledgements}

The authors are thankful to Antonio Maroto and Lev Kofman for useful
discussions.  The work of R.A was supported by
``Sonderforchschungsbereich 375 f$\ddot{\rm u}$r
Astro-Teilchenphysik'' der Deutschen Forschungsgemeinschaft.
 
%%%%%%%%%%%%%%%%%%%%%%%%%%%%%%%%%%%%%%%%%%%%%%%%%%%%%%%%%%%%%%%%%%%%%%%%%%%%%%%%%%%%%%%%

%\vspace*{-0.8truecm}

%%%%%%%%%%%%%%%%%%%%%%%%%%%%%%%%%%%%%%%%%%%%%%%%%%%%%%%%%%%%%%%%%%%%%%%%%%%%%%%%%%%%%%%%%%%%%

%%%%%%%%%%%%%%%%%%%%%%%%%%%%%%%%%%%%%%%%%%%%%%%%%%%%%%%%%%%%%%%%%%%%%%%%%%%%%%%%%%%%%%

\end{document}